\documentclass[lettersize,journal]{IEEEtran}
\usepackage{amsmath,amsfonts}
\usepackage{algorithm}
\usepackage[noend]{algpseudocode}
\usepackage{balance}
\usepackage{amssymb}
\usepackage{array}
\usepackage{textcomp}
\usepackage{stfloats}
\usepackage{url}
\usepackage{color}
\usepackage{verbatim}
\usepackage{booktabs}
\usepackage{cite}
\usepackage{enumerate}
\usepackage{graphicx}
\usepackage{subfigure}
\usepackage{filecontents}
\usepackage{multirow}
\usepackage{algpseudocode}

\begin{document}

\title{RIS-Assisted Beamfocusing in Near-Field IoT Communication Systems: A Transformer-Based Approach}

\author{Quan Zhou, Jingjing Zhao, Kaiquan Cai, and Yanbo Zhu\\
	\thanks{Q. Zhou, J. Zhao, K. Cai, and Y. Zhu are with the Beihang University, Beijing, China, and also with the State Key Laboratory of CNS/ATM, Beijing, China. (email:\{quanzhou, jingjingzhao, caikq, zhuyanbo\}@buaa.edu.cn). }
}

\markboth{}%
{Shell \MakeLowercase{\textit{et al.}}: A Sample Article Using IEEEtran.cls for IEEE Journals}


\maketitle
\def\mathbi#1{\textbf{\em #1}}
\begin{abstract}
The massive number of antennas in extremely large aperture array (ELAA) systems shifts the propagation regime of signals in internet of things (IoT) communication systems towards near-field spherical wave propagation. We  propose a reconfigurable intelligent surfaces (RIS)-assisted beamfocusing mechanism, where the design of the two-dimensional beam codebook that contains both the angular and distance domains is challenging. To address this issue, we introduce a novel Transformer-based two-stage beam training algorithm, which includes the coarse and fine search phases. The proposed mechanism provides a fine-grained codebook with enhanced spatial resolution, enabling precise beamfocusing. Specifically, in the first stage, the beam training is performed to estimate the approximate location of the device by using a simple codebook, determining whether it is within the beamfocusing range (BFR) or the none-beamfocusing range (NBFR). In the second stage, by using a more precise codebook, a fine-grained beam search strategy is conducted. Experimental results unveil that the precision of the RIS-assisted beamfocusing is greatly improved. The proposed method achieves beam selection accuracy up to 97\% at signal-to-noise ratio (SNR) of 20 dB, and improves 10\% to 50\% over the baseline method at different SNRs.
\end{abstract}


\section{Introduction}
\IEEEPARstart{R}{ecently}, the advent of sixth-generation wireless communication (6G) has catalyzed profound transformations across various industries, enhancing both industry operations and user experiences. To further improve system-level channel capacity, a novel approach was proposed to reconstruct the wireless channel by employing a virtual massive multiple-input multiple-output (MIMO) architecture formed by unmanned aerial vehicles (UAVs) \cite{gao2021JSAC}. These advancements have significantly increased the demand for high-speed, high-capacity, and low-latency communication solutions \cite{Garcia2021CST}. Specifically, the integration of massive MIMO and millimeter-wave technologies in 6G goes beyond simply increasing the number of antenna elements or expanding frequency bands; it facilitates a notable shift from far-field communication (FFC) to near-field communication (NFC) \cite{Bjornson2018TWC}. The Rayleigh distance is commonly employed to differentiate between near-field and far-field regions: areas within the Rayleigh distance are classified as near-field, while those beyond are categorized as far-field. Traditional antenna arrays, constrained by a small number of elements and lower frequency operations, exhibit limited near-field regions \cite{Cui2023TCM}. Conversely, the emerging 6G era, marked by extremely large antenna array (ELAA) and millimeter-wave frequencies, highlights the increasing importance of NFC \cite{Heath2016JSTSP}. For instance, when an antenna with 1.0-meter aperture and 60 GHz frequency, the near-field region extends to 400 meters, surpassing the typical 5G cell radius \cite{Pi2011CM, Rappaport2017TAP}. In the near-field, signal propagation becomes increasingly intricate, evolving from planar waves to spherical waveforms, which results in non-linear phase shifts across the antenna array, enabling a transition from beamsteering that considers only the angular domain to beamfocusing that takes into account both the angular and distance domains \cite{Marzetta2010TWC}.

Beamfocusing is particularly critical for the internet of things (IoT) network, which frequently operate in dense and highly variable propagation environments where traditional far-field communication models fail to provide consistent and efficient connections. As IoT systems continue to expand in deployment density and application diversity, these challenges underscore the necessity for advanced methodologies to ensure stable and efficient communication links. Whereas, traditional beam training methods oriented to beamsteering prove inadequate in near field, necessitating the development of beam training strategies oriented to beamfocusing \cite{Wang2023CST}. Promisingly, reconfigurable intelligent surfaces (RIS) can dynamically adjust the angle and phase of reflections to alleviate severe path loss and multipath fading in signal propagation, enhancing a more precised beamsteering. The energy efficiency and flexibility of RIS, combined with their ability to extend the effective near-field. In IoT communication systems, precise beamsteering and alignment are essential for ensuring high-quality communication. 

\subsection{Prior Works}
\subsubsection{Near-field wireless communications}
The transition from traditional massive MIMO to extremely large-scale MIMO (XL-MIMO) presents new challenges in near-field channel modeling, performance analysis, and channel estimation \cite{LuICC2021, LuTWC2022}. In far-field scenarios, the phase variation across antenna elements is linear under the uniform plane wave assumption. However, XL-MIMO-assisted near-field communication (NFC) systems introduce more complex requirements, such as the development of near-field beam codebooks, beam training, and channel estimation techniques. Novel transmission techniques, such as delay alignment modulation, leverage the ultra-spatial resolution of XL-MIMO to facilitate precise spatial multiplexing \cite{LiCL2022, LuLWC2022}. To establish a unified NFC model, Liu \textit{et al.} \cite{Liu2023OJCS} proposed an NFC performance framework, deriving signal-to-noise ratio (SNR) and power scaling laws for line-of-sight (LoS) conditions, which aid in optimizing XL-MIMO performance in near-field settings. Akrout \textit{et al.} \cite{AkroutICC2022} investigated mutual coupling effects in single-input single-output (SISO) NFC, considering finite antenna dimensions and minimal scattering, which are crucial for improving the accuracy of channel estimation in XL-MIMO systems. Xie \textit{et al.} \cite{XieLWC2023} introduced a codebook design tailored for NFC using uniform circular arrays (UCA), enhancing beamforming efficiency in NFC environments. These studies highlight the complexity of NFC performance and provide valuable insights for enhancing XL-MIMO systems. To analyze near-field performance, Nepa \textit{et al.} \cite{NepaMAP2017} introduced the principles of near-field beamfocusing (NFBF) in microwave antennas for short-range wireless systems, offering a performance evaluation framework based on key metrics such as the 3 dB focal point, focal shift, focusing gain, and sidelobe levels. Cui \textit{et al.} \cite{CuiMCOM2023} provided further insights into near-field boundaries, challenges, and potential applications, emphasizing the differences between far-field beamsteering and near-field beamfocusing. Wei \textit{et al.} \cite{WeiCL2022} addressed the challenges at the intersection of far-field and NFC in terahertz (THz) systems, focusing on channel models, estimation techniques, and hybrid beamforming methods for cross-domain communications.

\subsubsection{Beam training for NFC}
Near-field beam training requires searching across both angle and distance domains, making conventional single-beam training methods inefficient and error-prone. These challenges necessitate advanced strategies to fully exploit the potential of XL-MIMO systems. Recent studies have proposed various solutions to address these issues. The authors in \cite{BTNFC1} utilized the ``near-field rainbow" effect for broadband beam training, leveraging the spatial dispersion inherent in wideband signals to enable high-precision beam alignment. Zhou \textit{et al.} \cite{BTNFC2} introduced a two-stage multi-beam training method based on sparse linear arrays, which reduces the search space and enhances training efficiency. Simultaneously, deep neural networks (DNNs) have been explored to reduce overhead and improve coverage in complex propagation environments by enabling data-driven beam selection and real-time adaptation \cite{BTNFC3, BTNFC4}. To reduce the search complexity, multi-stage searching approaches have also proven effective, progressively narrowing the search space to enhance beam training accuracy while minimizing overhead \cite{BTNFC5, BTNFC6}. UCA-based codebook designs exploit spatial symmetry to simplify beam codebook construction, making them particularly suitable for near-field scenarios \cite{BTNFC7}. Besides, ``sense-then-train" frameworks combine environment sensing with beam training, allowing the system to dynamically adapt to varying user distributions and environmental conditions \cite{BTNFC8}. Whereas, these two-stage beam search algorithm suffers from an error propagation problem, wherein errors in the first stage cause the subsequent second stage becoming invalid or ineffective. Furthermore, Lu \textit{et al.} \cite{BTNFC9} proposed a multi-resolution codebook that balances the trade-off between training accuracy and complexity by employing hierarchical beamforming techniques. To make greater use of computing power, the authors in \cite{BTNFC10} designed a deep learning (DL)-based beam estimation method that leverages large datasets and computational power to predict optimal beam configurations, significantly reducing training time. Furthermore, conventional search algorithms may produce unexpected outcomes in the beam search task. In \cite{BTNFC12}, hash functions were used to index beams, enabling fast and scalable training for large antenna arrays. In addition to these methods, RIS-assisted communications offer promising solutions to mitigate near-field challenges \cite{BTNFC13, BTNFC14, BTNFC15}. However, these RIS-assisted algorithms that utilize RIS solely for the communication cascade and fail to fully leverage the potential of RIS.

{Based on the aforementioned related works, numerous existing studies have investigated NFC beam training and RIS-assisted communication. However, RIS has mainly been used as a passive relay for channel cascading, with its potential for beamfocusing underexplored. RIS can convert the far-field beam from the BS into a near-field beam, enhancing spatial resolution by controlling phase and amplitude. It also improves beamfocusing efficiency by optimizing both angular and distance domains, thereby reducing interference. As a passive beamforming approach, RIS ensures energy efficiency and scalability, making it ideal for large-scale IoT systems.}

\subsection{Motivations and Contributions}
ELAA effectively extends transmission distances and mitigates path loss, addressing high-frequency attenuation in next-generation mobile communications. In fact, within this extended range, only about one-tenth of the Rayleigh length can implement beamfocusing, while the remainder relies on beamsteering \cite{Liu2023OJCS}. Accordingly, the near-field can be divided into beamfocusing range (BFR) and none-beamfocusing range (NBFR). Besides, as the size of beam codebook increases, traditional exhaustive search methods become computationally expensive and impractical for NFC systems. RIS-assisted communication offers a promising approach to address these challenges by enabling flexible beam coverage. Additionally, traditional beam training methods for beam selection based on classification models and cross-entropy loss function incurs computation overhead when the explosive growth of codes in the codebook. In contrast, mean square error (MSE) loss function are well-suited for regression tasks, modelling continuous spatial relationships by capturing fine-grained variations \cite{Bishop2006Springer}. In fact, when dealing with continuous output variables and assuming that the input variables or noise follow a Gaussian distribution, MSE is superior to the cross-entropy \cite{Goodfellow2016MITPress}. For the beam search task, this is not a classification issue but a detection problem that involves predicting continuous beam codes. Accordingly, MSE is a promising solution to turn large-scale classification issues into detection tasks.

To address these challenges, this paper proposes a Transformer-based beam training framework for the RIS-assisted NFC systems. The primary contributions are summarized as follows:
\begin{itemize}	
	\item We propose a RIS-assisted near-field communication architecture that improves the beamfocusing range of the base station (BS) and significantly extends the beam training process. The proposed approach overcomes the limitations of traditional beamforming techniques, which are ineffective in near-field environments, particularly in IoT communication scenarios involving ELAA. 
	\item We introduce an improved Transformer-based beam training algorithm, which enables both coarse and fine beam searching phase. By transforming the beam code selection task into a device position detection and location-mapping-beamcode process, our model efficiently navigates the two-dimensional search space, including angular and distance domains, thereby improving beamforming accuracy and adaptability in dynamic environments.
	\item Numerical results demonstrate that RIS-assisted near-field beamfocusing effectively enhances array gain, achieving a promising beam selection accuracy and beamfocusing gains. In addition, the proposed method provides significant improvements compared with baseline methods. The two-stage process reduces the beam search overhead and complexity and is well suited for large-scale near-field deployments.
\end{itemize}

\subsection{Organization}
The remainder of this paper is organized as follows: Section II formulates the system model and problem. Section III describes the proposed beam training algorithm. In Section IV, we present simulation results and performance analysis. Finally, Section V concludes the paper and discusses future research directions. 

{\textit{Notations:} Boldface symbols denote vectors (lower case, e.g., $\mathbf{a}$) and matrices (upper case, e.g., $\mathbf{A}$). The transpose and conjugate transpose (Hermitian) operations are represented by $(\cdot)^{T}$ and $(\cdot)^{H}$, respectively. The set of $P \times Q$-dimensional complex and real matrices is denoted by $\mathbb{C}^{P \times Q}$ and $\mathbb{R}^{P \times Q}$. For indexing, $a[p]$ refers to the $p$-th entry of vector $\mathbf{a}$, while $\mathbf{A}[p, q]$ denotes the entry in the $p$-th row and $q$-th column of matrix $\mathbf{A}$. The real part of vector $\mathbf{a}$ is written as $\operatorname{Re}\{\mathbf{a}\}$, and the trace of matrix $\mathbf{A}$ is $\operatorname{Tr}(\mathbf{A})$. The operator $\operatorname{diag}(\mathbf{a})$ generates a diagonal matrix with elements of $\mathbf{a}$ on the main diagonal, whereas $\operatorname{diag}(\mathbf{A})$ extracts the main diagonal of $\mathbf{A}$ as a column vector. A positive semi-definite matrix $\mathbf{A}$ is indicated by $\mathbf{A} \succeq 0$. The $K$-dimensional identity matrix is $\mathbf{I}_K$. Norms are defined as follows: the 2-norm of vector $\mathbf{a}$ is $\|\mathbf{a}\|_2$, and the spectral and Frobenius norms of matrix $\mathbf{A}$ are $\|\mathbf{A}\|_2$ and $\|\mathbf{A}\|_F$, respectively. The rank of $\mathbf{A}$ is $\operatorname{rank}(\mathbf{A})$. For a complex number $a$, its magnitude and phase are denoted by $|a|$ and $\angle a$, respectively.}

\begin{figure}[tp]
	\centering
	\includegraphics[scale=0.13]{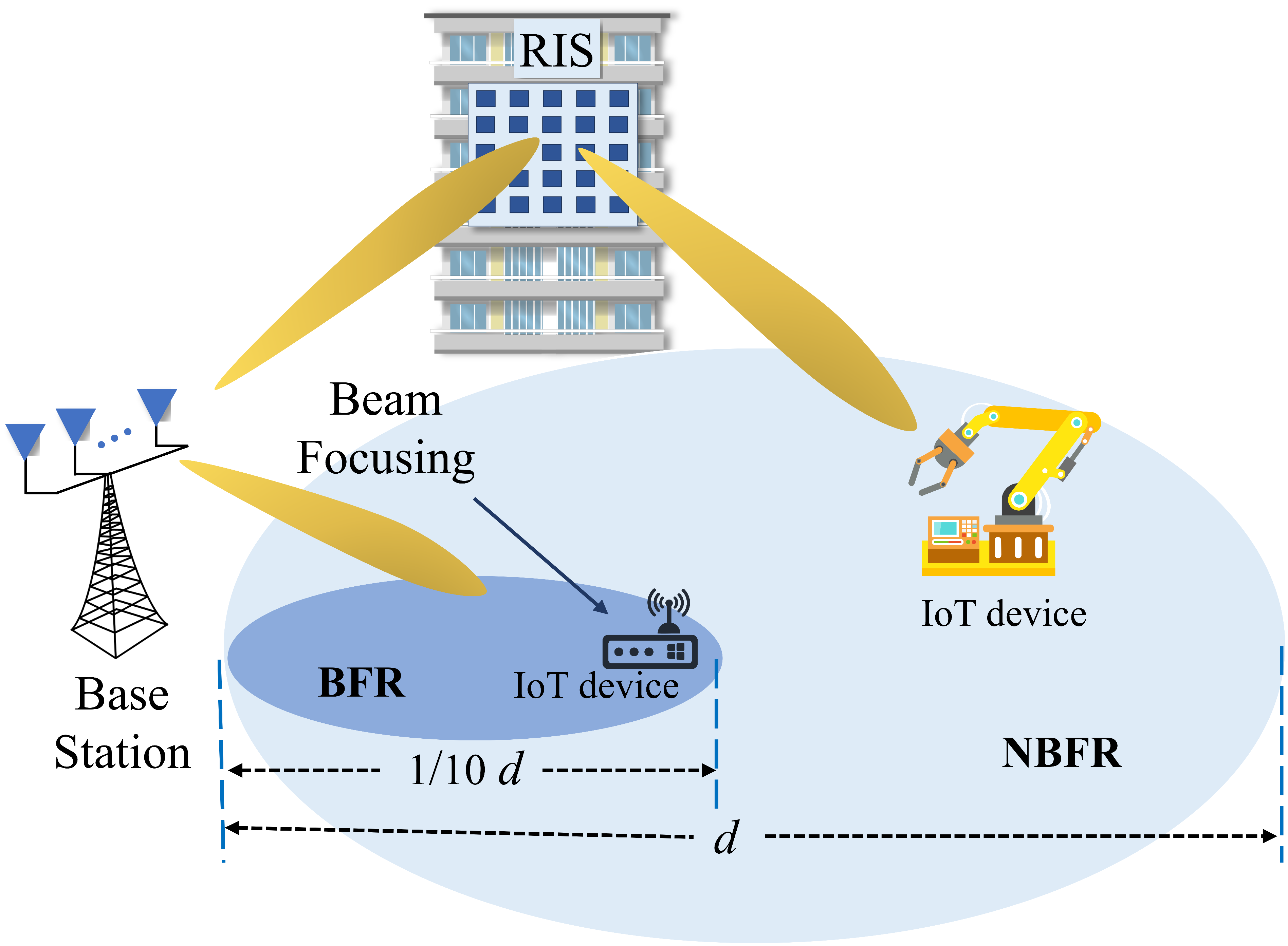}
	\caption{Illustration of RIS-assisted near-field beamfocusing between the BS and device.}
	\label{nearfield} 
\end{figure} 
\section{System Model and Problem Formulation}
\subsection{System Model}
As illustrated in Fig. \ref{nearfield}, we consider a downlink communication system. The BS is equipped with a uniform linear array (ULA) comprising $N$ elements. The RIS is deployed with a uniform planar array (UPA) consisting of $M$ passive reflecting elements. The device is configured with a single antenna. The distance between BS and RIS is smaller than the BS's Rayleigh distance $d = 2D_{\text{BS}}^2/\lambda$, where $D_{\text{BS}}$ denotes the size of the BS aperture and $\lambda$ is the wavelength. By assessing whether the device is within the BFR or NBFR, the BS establishes the primary communication channel, either with direct assistance from the BS or through the RIS by adjusting the beamforming matrix.

\subsubsection{BS-to-RIS channel}
The channel between the BS and the RIS is expressed as:
\begin{equation}
	\textbf{H}_{\text{BR}} = [\mathbf{h}_{\text{BR}}^1, \cdots, \mathbf{h}_{\text{BR}}^{m}, \cdots, \mathbf{h}_{\text{BR}}^M]^T,
\end{equation}
where \( \mathbf{h}_{\text{BR}}^{m} \) is
\begin{equation}
	\mathbf{h}_{\text{BR}}^{m} = \left[k_{m,1}e^{-j\frac{2\pi}{\lambda_c}d_{m,1}}, \cdots, k_{m,N}e^{-j\frac{2\pi}{\lambda_c}d_{m,N}}\right],
\end{equation}
where $k_{m, n}$ represents the free-space path loss between the BS's $n$-th element and the RIS's $m$-th element, $\lambda_c$ is the carrier wavelength, and $d_{m, n}$ is the distance between them, introducing a phase shift proportional to $e^{-j\frac{2\pi}{\lambda_c}d_{m,n}}$.

\subsubsection{BS-to-Device channel}
The channel between the BS and the device is formulated as:
\begin{equation}
	\begin{aligned}
		\mathbf{h}_{\text{BU}} = \left[k_{1}e^{-j\frac{2\pi}{\lambda_c}d_{1}}, \cdots, k_{N}e^{-j\frac{2\pi}{\lambda_c}d_{N}}\right],
	\end{aligned}
\end{equation}
where $k_n$ is the path loss and $d_n$ is the distance between the BS's $n$-th element and the device, introducing a phase shift of $e^{-j\frac{2\pi}{\lambda_c}d_n}$.

\subsubsection{RIS-to-Device channel}
The channel between RIS and device is given by:
\begin{equation}
	\begin{aligned}
		&\mathbf{h}_{\text{RE}} = \left[k_{1}e^{-j\frac{2\pi}{\lambda_c}d_{1}}, \cdots, k_{m}e^{-j\frac{2\pi}{\lambda_c}d_{m}}, \cdots, k_{M}e^{-j\frac{2\pi}{\lambda_c}d_{M}}\right],
	\end{aligned}
\end{equation}
where \( k_{m} \) is the path loss between the RIS's \( m \)-th element and the device, \( d_{m} \) is the distance between the RIS's \( m \)-th element and the device. 

\subsection{Problem Formulation}
To achieve near-field beamfocusing, we utilize the RIS to enhance channel gains and improve the overall signal strength. The received signal is expressed in two forms based on the distance between the device and the BS. The received signal $\mathcal{Y}$ is expressed as:
\begin{equation}
	\mathcal{Y} = 	\left\{
	\begin{aligned}
		 & \mathbf{h}^{H}_{\text{RE}} \mathbf{F}x + u, & 0 < r \leq d/10, \\		
		 & \mathbf{h}^{H}_{\text{RE}}\mathbf{\Phi}\mathbf{H}_{\text{BR}}\mathbf{F}x + u, & d/10 < r \leq d, 
	\end{aligned}
	\right.
\end{equation}
where $u\sim\mathcal{CN}(0, \sigma^2)$ denotes additive white Gaussian noise (AWGN), $x$ is the transmitted signal from BS, and $\mathbf{F}$ represents the beamforming matrix, $\boldsymbol{\Phi} = \mathbf{diag}\left(e^{j\theta_1}, \ldots, e^{j\theta_M}\right)$ is the phase shift matrix of the RIS, and $\theta_m \in [0, 2\pi)$. {$d$ is the Rayleigh distance for near field, which characterizes the coverage of the beamfocusing \cite{Liu2023OJCS}.} $r$ indicts the distance between the BS and the device, and if $r \leq d/10$, the direct BS-to-device channel takes on a primary role; otherwise, the cascaded channel is employed. Accordingly, the communication rate $R$ is given by:
\begin{equation}
	R = \left\{
		\begin{aligned}
		 & \log_2\left(1+\frac{\left|\mathbf{h}^{H}_{\text{RE}} \mathbf{F}\right|^{2}}{\sigma_{u}^{2}}\right), & 0 < r \leq d/10, \\		
		 &\log_2\left(1+\frac{\left|\mathbf{h}^{H}_{\text{RE}}\mathbf{\Phi}^{H}\mathbf{H}_{\text{BR}}{\mathbf{F}}\right|^{2}}{\sigma_{u}^{2}}\right), & d/10 < r \leq d.
		\end{aligned}
		\right.
\end{equation}

We introduce the array gain $\zeta$, which quantifies the impact of beamforming on the received signal strength at the device. $\zeta$ reflects the beamforming efficiency and is determined by the array's directional properties, which is formulated as:
\begin{equation}
	\zeta = \frac{1}{N}\left| \mathbf{a}^{H}(\theta, r) \mathbf{a}(\theta, r) \right|,
\end{equation}
where $\mathbf{a}(\theta, r)$ is the array steering vector for the device at angle $\theta$ and distance $r$. {The practical significance of $\zeta$ lies in its determination of the beamfocusing performance. A smaller focal depth $d_f = r_{max} - r_{min}$, which represents the effective coverage of the beam over the target area, indicates superior beamfocusing capabilities. This leads to enhanced signal quality and minimized interference when the device operates outside this range. Additionally, the constraint on the 3 dB main lobe width of $\mathbf{a}(\theta, r)$ ensures that the beam energy is efficiently concentrated in the desired direction and at the intended distance.} The optimization problem is formulated as:
\begin{equation}\label{eq:argmax_problem1}
\mathbf{F}_{\tilde{l}, \tilde{s}} = \left\{
	\begin{aligned}
		&\underset {\mathbf{F}_{l, s}} { \operatorname {arg\,max} } \, \log_2\left(1+\frac{P_{t}\left|{\mathbf{h}^{H}_{\text{RE}}\mathbf{F}_{l, s}}\right|^{2}}{\sigma_{u}^{2}}\right),\\
		&\underset {\mathbf{F}_{l, s}, \mathbf{\Phi}} { \operatorname {arg\,max} } \, \log_2\left(1+\frac{P_{t}\left|\mathbf{h}_{\text{RE}}^{H}\mathbf{\Phi}^{H}\mathbf{H}_{\text{BR}}{\mathbf{F}_{l, s}}\right|^{2}}{\sigma_{u}^{2}}\right),
		\end{aligned}
		\right.
\end{equation}
where $P_{t}$ represents the transmit power, $\mathbf{F}_{l, s}$ is the precoding matrix at the BS, which is designed to adapt the transmitted signals, while $\mathbf{\Phi}$ adjusts the RIS's phase shifts to enhance the signal strength received by the device. The optimization aims to maximize the data rate for BS-directed or RIS-assisted communication depending on the device’s location.

{According to \cite{Cui2023TCM}, the sampling constraint for the near-field can be expressed as:}
\begin{equation}
|r_{l} - r_{l-1}| \geq 2\lambda \beta_{\Delta}^2 |r_l r_{l-1}|/d^2_e, 
\end{equation}
where $\beta_{\Delta}$ accounts for the beamforming pattern's spatial variation, and $d_e$ represents the element spacing in the antenna array. Consequently, the sampling distance is expressed as $r(\theta, \beta) = d^2/(2s\lambda \beta_{\Delta}^2)$, where $s$ is the sampling index. In traditional near-field systems, the sampling distance is constrained by the physical arrangement of antenna elements. However, by incorporating a RIS with a larger effective aperture, $D_{\text{RIS}} = Md_e$, the range of near-field $d = 2(Md_e)^2/\lambda$ becomes adjustable, allowing for more flexible and equidistant sampling. Therefore, equidistant sampling is defined as $|r_l - r_{l-1}| = \Delta r$, where $ \Delta r$ denotes the fixed sampling interval. This flexibility facilitates the design of the near-field codebook:
\begin{equation}
	\mathcal{C}_{\text{RIS}} = \left\{c_{i, j} \,|\, i = 1, 2, \ldots, I, j = 1, 2, \ldots, J\right\},
\end{equation}
where $\mathcal{C}_{\text{RIS}}$ denotes the codebook for the RIS-assisted approach, $c_{i, j}$ is $(i, j)$-th the beam code, and $ I\times J $ is the total number of sampling points. By adopting the RIS-assisted approach, beamfocusing can be achieved within the BS's Rayleigh distance.

\section{Two Stage Beam Training Scheme}
Leveraging the near-field mechanisms and the flexibility of the RIS, we propose a two-stage beam training scheme. As shown in Fig. \ref{fig:twostage}, in the first stage, coarse search phase is performed by leveraging the simple codebook, estimating the device’s location within the BFR or NBFR. The second stage involves fine search phase with a more precise codebook for beam training. If the device within the BFR, BS-based beam training is employed. Otherwise, RIS-assisted beam training is implemented.

\subsection{Codebook Design}
To facilitate coordination between the BS and RIS, we introduce two reference coordinate systems: one centered at the BS and the other at the RIS. For the BS-directed beam training approach, the codebook for the coarse searching phase consists of a $2 \times 2$ grid, with four codewords to reduce false positives compared to two. For the RIS-assisted communication, the codebook is refined to an $L \times S$ grid, where $L$ and $S$ denote the total number of divisions for the angular and distance domains, providing a finer resolution for beam training. This expanded codebook is then used for estimating the device's position and refining the beamforming process. This process incorporates $\mathcal{Y}$ the feed back from the device. The optimal beam index $(\tilde{l}, \tilde{s})$ in $\mathcal{Y}$ is defined as:
\begin{equation}
\mathcal{Y}_{(\tilde{l}, \tilde{s})} = \max \left\{ \mathcal{Y}_{(1, 1)}, \dots, \mathcal{Y}_{(L, S)} \right\}.
\end{equation}
Accordingly, it is means that beamforming according to the beam index $(\tilde{l}, \tilde{s})$ yields the maximum $\zeta$. Based on the two reference coordinate systems, we define the two-stage beam training mechanisms comprising the coarse searching phase and the fine searching phase, which are described as follows:

\begin{figure}[tp]
	\centering
	\subfigure[Coarse searching]{
		\includegraphics[width=0.18\textwidth]{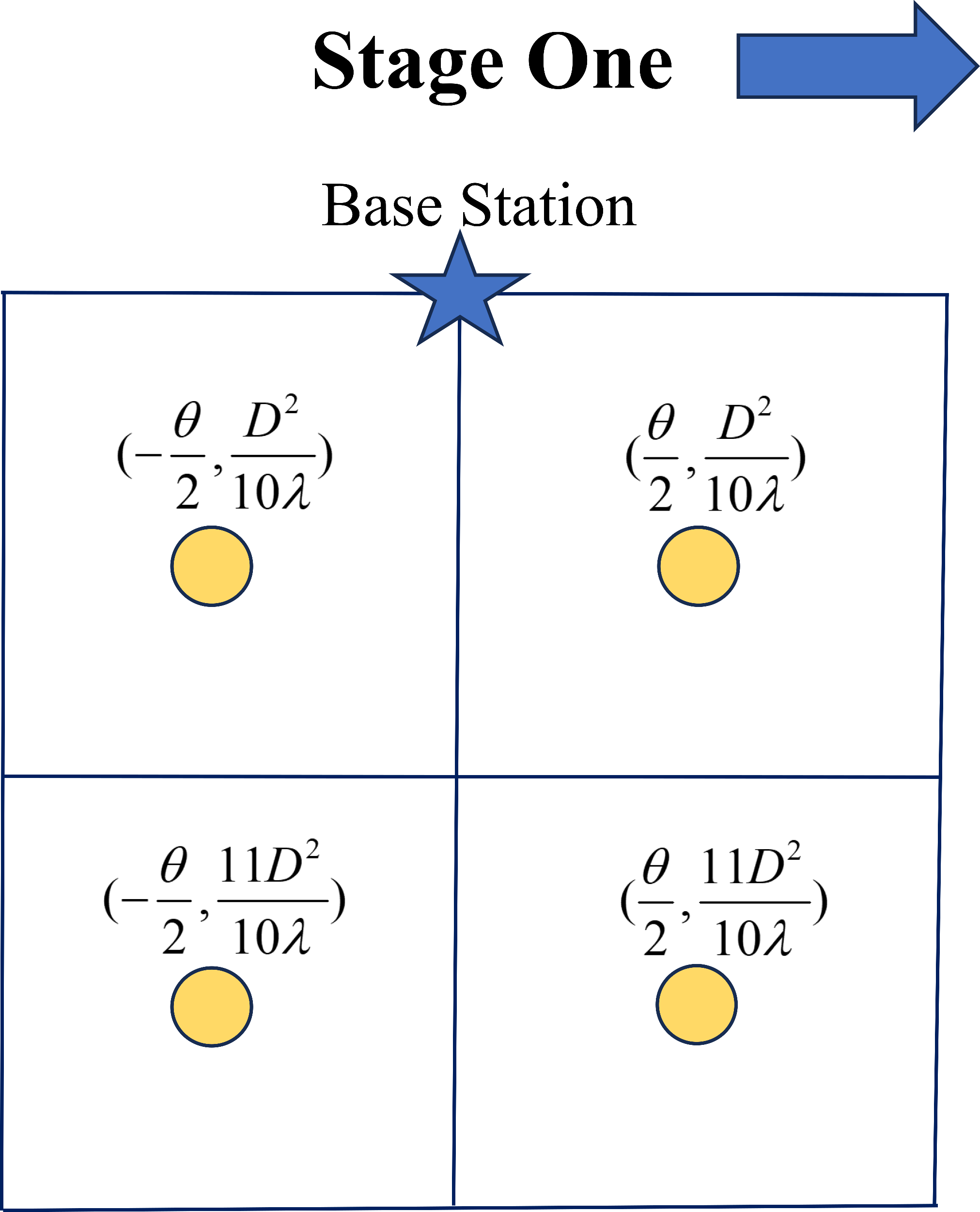} 
	}
	\subfigure[Fine searching]{
		\includegraphics[width=0.25\textwidth]{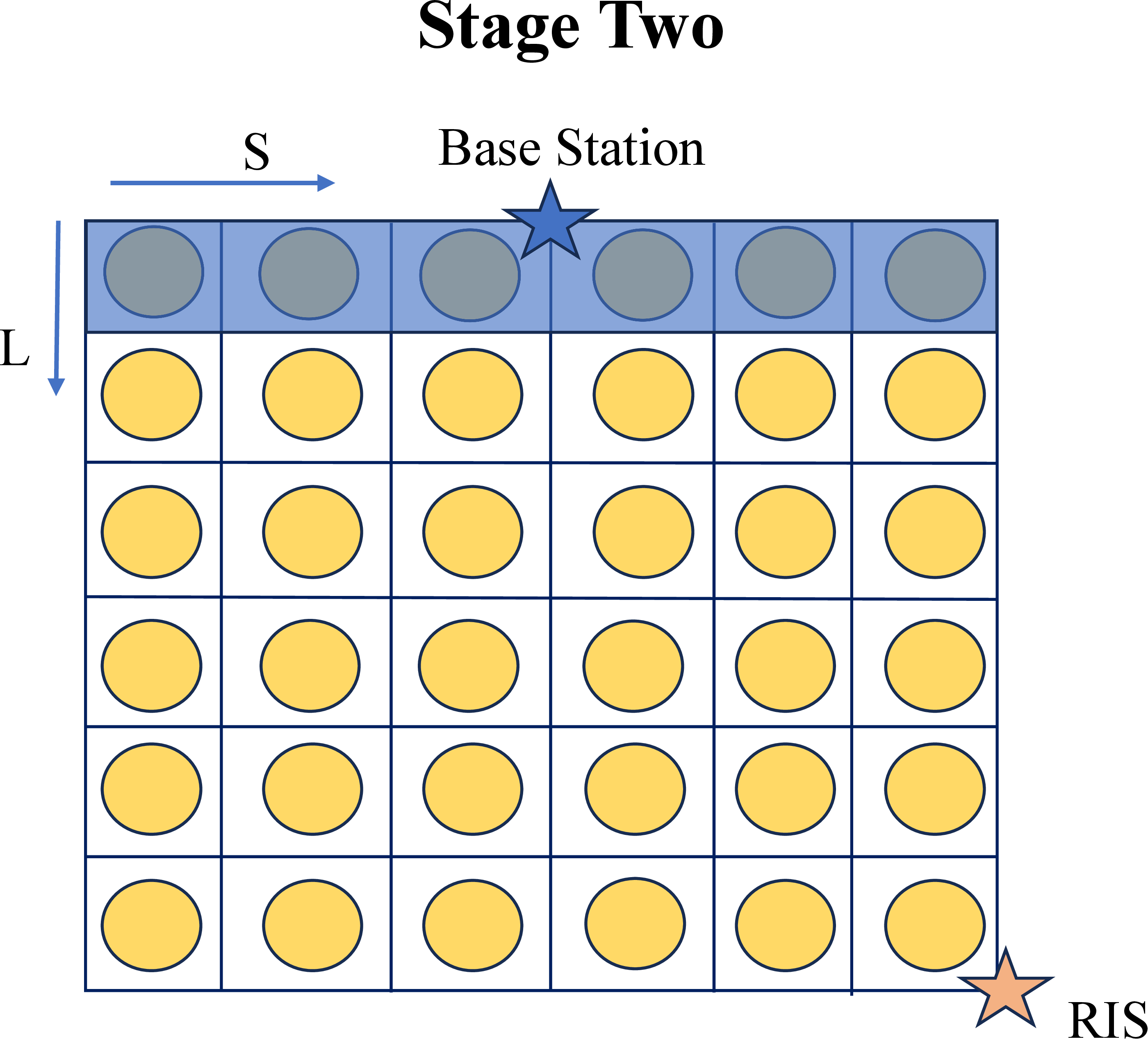}
	}
	\caption{The proposed two-stage codebook for near-field beam training.}
	\label{fig:twostage}
\end{figure}

\subsubsection{Coarse searching} In the coarse searching phase, the $2\times2$ codebook is designed in conjunction with a location prediction model to estimate the position of the device and determine whether it lies within the BFR or the NBFR. As illustrated in Fig. \ref{fig:twostage} (a), the coarse searching codebook consists of four codes, which are formulated as:
\begin{equation}
		\mathcal{C}_{\text{BS}} = 
\begin{aligned}
	\left\{ \begin{array}{c}
		c_1\left(-\frac{\theta}{2},\frac{11D^2}{10\lambda}\right),\ c_2 \left(\frac{\theta}{2}, \frac{11D^2}{10\lambda}\right) \\
		
		c_3\left(-\frac{\theta}{2},\frac{D^2}{10\lambda}\right),\  c_4\left(\frac{\theta}{2}, \frac{D^2}{10\lambda}\right)
	\end{array} \right\},
\end{aligned}
\end{equation}
where $D$ denotes the size of the BS aperture, $\theta$ represents the range of beam angular direction. $\{c_1, c_2, c_3, c_4\}$ represent different searching positions relative to the BS. After estimating the location of the device and mapping into $\mathcal{C}_{\text{BS}}$, the system transitions from coarse searching to fine searching, using a more precise codebook for finer beamfocusing. This involves applying a distance and angle prediction algorithm to identify the device’s location for either BS-directed or RIS-assisted beam training. 

\subsubsection{Fine searching} To enable the beam codes to accommodate dense devices when performing beam assignment, the following codebook sampling principle is defined as:
\begin{equation}
	\left\{
	\begin{aligned}
		&\phi_l = \frac{2l-L-1}{L},& l = 1, 2, \cdots, L, \\
		&r_{l,s}=\frac{ds}{S},& s = 1, 2, \cdots, S,
	\end{aligned}
	\right.
\end{equation}
where $\phi_n$ denotes the real angles of the device with respect to the BS. Assuming the device’s relative angles and distances with respect to the BS in the near field are denoted as $(r_l, \theta_l)$, and the relative parameters between the RIS and BS are $(r_s, \theta_s)$, the relative angles and distances between the device and the RIS are calculated as:
\begin{equation}
	\left\{
	\begin{aligned}
		&r_{\text{ris}} = \sqrt{r_{s}^2+r_{l}^2-2 r_{s} r_{l} \cos(\theta_{s}-\theta_{l})}, \\
		&\theta_{\text{ris}}=\arctan\left(\frac{r_{l}\sin(\theta_{l})-r_{s} \sin(\theta_{s})}{r_{s}\cos(\theta_{l})-r_{s} \cos(\theta_{s})}\right)+\theta_{l}.
	\end{aligned}
	\right.
	\label{coordinateconvert}
\end{equation}
Thus, as shown in Fig. \ref{fig:twostage} (b), both the BS-directed and RIS-assisted approaches are considered in the beam training process.
\begin{itemize}
	\item BS-directed: When the first phase of beam searching indicates that the device is within the BFR, the BS performs near-field beam searching. The uplink feedback from the device is used to refine its location, enabling beamfocusing.
	\item RIS-assisted: Upon the device is detected within the NFR, the codebook is converted into coordinates relative to the RIS. Subsequently, near-field beam searching is conducted through the RIS, utilizing uplink feedback to ascertain the device's position, followed by beamfocusing based on the cascaded channel.
\end{itemize}

\subsection{Transformer-Based Beam Training for NFC}
\renewcommand{\arraystretch}{1.2} 
\begin{table}[tp]
	\caption{The details of each layer in the proposed near-field beam training architecture.}
	\label{tab:Margin_settings}
	\fontsize{8}{10}\selectfont  
	\centering
	\begin{tabular}{|>{\raggedright\arraybackslash}p{2cm}|>{\raggedright\arraybackslash}p{3.5cm}|>{\raggedright\arraybackslash}p{2.2cm}|}
		\hline
		\textbf{Module} & \textbf{Layer} & \textbf{Output} \\
		\hline
		Input Layer & \centering{/}& 100 $\times$ 100 \\
		\hline
		\multirow{8}{*}{Input  Embedding} & Conv1D(filters 64, size 1$\times$3, padding 1)+BN+Relu+ MaxPool1D(size 2, stride 2) & \multirow{3}{*}{50 $\times$ 50} \\
		\cline{2-3}
		& Conv1D(filters 64, size 1$\times$3, padding 1)+BN+Relu+ MaxPool1D(size 2, stride 2) & \multirow{3}{*}{25 $\times$ 25}\\
		\cline{2-3}
		& Conv1D(filters 64, size 1$\times$3, padding 1)+BN+Relu+ MaxPool1D(size 2, stride 2) & \multirow{3}{*}{13 $\times$ 13} \\
		\hline
		\multirow{12}{*}{Attention Module} & Multi-Head Attention & $13 \times 13$ \\
		\cline{2-3}
		& Feed Forward + Add \& Norm& $13 \times 13$ \\
		\cline{2-3}
		& Add \& Norm & $13 \times 13$ \\
		\cline{2-3}
		& Multi-Head Attention & $13 \times 13$ \\
		\cline{2-3}
		& Add \& Norm & $13 \times 13$ \\
		\cline{2-3}
		& Feed Forward & $13 \times 13$ \\
		\cline{2-3}
		& Add \& Norm & $13 \times 13$ \\	
		\cline{2-3}
		& Masked Multi-Head Attention & $13 \times 13$ \\
		\cline{2-3}
		& Add \& Norm & $13 \times 13$ \\
		\cline{2-3}
		& Feed Forward + Add \& Norm & $13 \times 13$ \\
		\hline
		\multirow{3}{*}{Position Detector} & Flatten & (None, 169) \\
		\cline{2-3}
		& Dense+Relu & (169, 128) \\
		\cline{2-3}
		& Dense+Sigmoid & (128, 2) \\
		\hline
	\end{tabular}
\end{table}

\begin{figure*}[tp]
	\centering
	\includegraphics[scale=0.24]{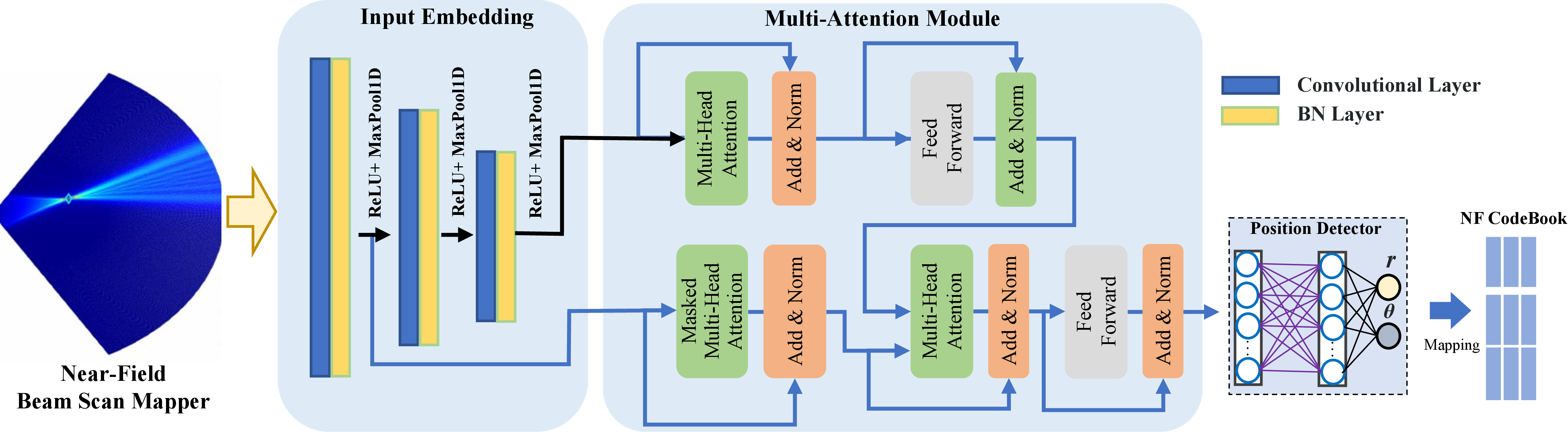}
	\caption{The proposed Transformer-based architecture for near-field beam training.}
	\label{TransformerModel} 
\end{figure*} 

In DL-based beam training approach, collecting a large amount of data is challenging. Consequently, traditional DL algorithms are not directly applicable to communication scenarios. Different from convolutional neural networks (CNNs), which specialize in local feature extraction, the Transformer architecture prioritizes global information, showcasing significant potential for handling complex tasks. The structure of each layer in the proposed architecture is detailed in \textbf{Table \ref{tab:Margin_settings}}, and the main modules are elaborated as follows:
\subsubsection{Input embedding} {The inputs are beam scan maps that are divided into several patches, which are subsequently flattened and embedded through a trainable linear transformer before being passed into the Transformer. To optimize data processing within the proposed model, the beam maps $\mathbf{M}=[H, W, C]$ are partitioned into $N = HW/P^2 \in \mathbb{R}^{P\times P\times C}$. Subsequently, each patch is flattened and transformed into a high-dimensional vector $\textbf{z}_i$ by a trainable linear transformation matrix $\mathbf{E}$. This transformation process is mathematically expressed as:}
	\begin{equation}
	\textbf{z}_i = \textbf{x}_i \mathbf{E} + b,
\end{equation}
where $\textbf{x}_i$ is the flattened $i$-th patch, $b$ represents bias. These embeddings serve as input to the Transformer model.

\subsubsection{Multi-head self-attention mechanism} The multi-head self-attention mechanism, serving as the core module of the Transformer architecture, captures similarity relationships among input patches. Transformer models employ three trainable projection matrices to generate the Query (\textbf{Q}), Key (\textbf{K}), and Value (\textbf{V}) vectors from the input embeddings: $\mathbf{W}^Q \in \mathbb{R}^{D \times d_k}$ is the matrix for projecting the input into the Query space; $\mathbf{W}^K \in \mathbb{R}^{D \times d_k}$ is the matrix for projecting the input into the Key space; $\mathbf{W}^V \in \mathbb{R}^{D \times d_v}$ is the matrix for projecting the input into the Value space. The computation of $\mathbf{Q}^i$, $\mathbf{K}^i$, and $\mathbf{V}^i$ is performed using matrix multiplication as follows:
\begin{equation}
\mathbf{Q}^i = \mathbf{z}_i \mathbf{W}^Q, \quad \mathbf{K}^i = \mathbf{z}_i \mathbf{W}^K, \quad \mathbf{V}^i = \mathbf{z}_i \mathbf{W}^V,
\end{equation}
where $\mathbf{Q}^i, \mathbf{K}^i \in \mathbb{R}^{N \times d_k}$ are used to calculate the attention scores; $\mathbf{V}^i \in \mathbb{R}^{N \times d_v}$ is the Value matrix, which represents the content information to be passed forward. $\mathbf{W}^Q$, $\mathbf{W}^K$, and $\mathbf{W}^V$ are trainable projection matrices learned during the model's training phase. These projections enable the model to transform the input into three different vector spaces, which are then used in the self-attention mechanism to calculate attention weights and process information efficiently. These weights reflect the importance relationships between patches, which can be expressed as:
{\begin{equation}
		f_{\textbf{Att}}(\textbf{Q}^{i},\textbf{K}^{i}, \textbf{V}^{i}) = \textbf{Softmax}\left(\frac{\textbf{Q}^{i}\textbf{K}_{i}^T}{\sqrt{d_k}}\right)\textbf{V}_{i}.
\end{equation}}
The multi-head attention mechanism performs parallel computations across multiple independent attention heads. {The value of $\textbf{head}_{i}$ refers to an individual attention mechanism operating on different parts of the input, enabling the model to capture diverse aspects of the input data.} Assume that there are $h$ heads, each with separate $\textbf{Q}^i$, $\textbf{K}^i$, $\textbf{V}^i$ matrices, and the multi-head attention operation is represented as:
\begin{equation}
	{\begin{aligned}
		&f_{\textbf{MH}}{(\textbf{Q}, \textbf{K}, \textbf{V})} = \mathbb{C}\left(\textbf{head}_1, \textbf{head}_2, \cdots, \textbf{head}_h\right)\textbf{W}^{O},\\
		&\textbf{head}_{i} = f_{\textbf{Att}}\left(\textbf{Q}^{i}\textbf{W}_i^{Q}, \textbf{K}^{i}\textbf{W}_i^{K}, \textbf{V}^{i}\textbf{W}_i^{V}\right),
	\end{aligned}}
\end{equation}
\noindent where $\mathbb{C}\left(\cdot\right)$ denotes concatenation of the outputs from all heads, $\textbf{W}^{O}$ is the output matrix, $\textbf{W}_i^{Q}, \textbf{W}_i^{K}, \textbf{W}_i^{V}$ are the projection matrices. {$f_{\textbf{MH}}{(\textbf{Q}, \textbf{K}, \textbf{V})}$ enables the process of calculating the attention weights and applying them to the input to produce the attended output.} The outputs of all heads are concatenated and transformed through a linear layer to integrate information from different features.
\begin{algorithm}[tp]
	\caption{Cascade Channel Generation for RIS-assisted Beamfocusing}
	\begin{algorithmic}[1]
		\State \textbf{Initialize}: the codebook size $S$ and $L$, the total number of datasets $N_{c}$.
		\State \textbf{Generate Cascade Channel} $G_{ris}(N, M, N_c)$:
		\While{\textbf{True}}
		\State Random sample angles and distances: $\phi_{br}$, $\phi_{ru}$, $r_{br}$, $r_{ru}$;
		\State Channel gains: $\{G_{ris}, \phi_{ru}, r_{ru}, m, n\}$;
		\State Initialize $h_{br}$ and $h_{ru}$;
		\For{$n_c = 1$ to $N_c$}
		\For{$n = 0$ to $N-1$}
		\For{$m = 0$ to $M-1$}
		\State \textbf{Update} $h_{br}(n, m) = k_{n, m}e^{-j\frac{2\pi}{\lambda_c}d_{n, m}}$;
		\EndFor
		\EndFor
		\For{$m = 0$ to $M-1$}
		\State \textbf{Update} $h_{ru}(m, 1) = k_{m,1}e^{-j\frac{2\pi}{\lambda_c}d_{m, 1}}$;
		\EndFor
		\EndFor
		\State \textbf{Compute} $G_{ris} = h_{br}\mathbf{\Phi} h_{ru}$;
		\EndWhile
		\State \textbf{Generate Codebook} $\mathcal{C}(L, S)$:
		\For{$s = 1$ to $S$}
		\For{$l = 1$ to $L$}
		\State $\theta = \frac{2l - L - 1}{L}$, $r_s = (d/S)s$;
		\State $c(l, s) = \frac{1}{L}e^{\left(-j \frac{2\pi}{\lambda_c}(ld\sin(\theta)-\delta_n)\right)}$.
		\EndFor
		\EndFor
	\end{algorithmic}
	\label{algorithm1}
\end{algorithm}

\subsubsection{Masked multi-head attention} In Transformer module, the padding mask and sequence mask are essential for handling diverse tasks and requirements. The padding mask specifically eliminates the influence of padding positions, which are introduced to standardize sequence lengths during batch processing, ensuring that these positions do not affect attention computations. The padding mask is formulated as:
 \begin{equation}
 \textbf{Mask}_p(i, j) = \left\{
	\begin{aligned}
		&0,\ \text{if}\ \text{position}\ j\ \text{is}\ \text{padding},\\
		&1,\ \text{otherwise}. \\
	\end{aligned}
	\right.
\end{equation}
Moreover, the sequence mask ensures that previous time steps are considered when predicting the current time step, preventing the model from accessing future positions and eliminating the risk of information leakage. The sequence mask is formulated as:
  \begin{equation}
		\textbf{Mask}_s(i, j) = \left\{
		\begin{aligned}
			&0,\ \text{if}\ j\ \textgreater\ i,\\
			&1,\ \text{otherwise}. \\
		\end{aligned}
		\right.
	\end{equation}
Accordingly, self-attention with mask is incorporated into the scaled dot-product attention calculation to mask out unnecessary parts, which can be expressed as:
{\begin{equation}
		f_{\textbf{MaskAtt}}\left(\textbf{Q}^{i}, \textbf{K}^{i}, \textbf{V}^{i}\right) = \textbf{Softmax}\left(\frac{\textbf{Q}^{i}(\textbf{K}^{i})^T}{\sqrt{d_k}}+\textbf{Mask}\right)\textbf{V}^{i}.
\end{equation}
The masked multi-head attention mechanism $f_{\textbf{MaskAtt}}$ allows Transformer modules to effectively handle sequences of varying lengths and prevents information leakage in auto-regressive tasks. The padding mask is applied to eliminate the influence of padding positions during batch processing, ensuring accurate attention computation. Additionally, the sequence mask prevents future position information leakage during training, thereby improving robustness and prediction accuracy.} By incorporating this masking strategy, the model can optimize beam selection while ensuring that past beamforming decisions do not interfere with future predictions, ultimately improving the efficiency of the communication system.

{In contrast to CNNs and Long short-term memory (LSTM), which process features either locally or sequentially, the Transformer's self-attention mechanism enables the computation of global interactions across all input patches simultaneously. This parallelized global processing capability is particularly essential for near-field beam training, where the optimal selection of codewords relies on capturing joint angular-distance correlations spanning the entire spatial domain. By eliminating the error propagation commonly associated with multi-stage search frameworks, our Transformer-based approach achieves a direct mapping from user uplink feedback to beam indices with sub-wavelength precision.}\\

\subsubsection{Loss function} 
The MSE loss function can effectively reduce prediction errors and improve positioning accuracy by minimizing the squared difference between predicted and actual values. The MSE loss function offers a stable error metric that improves the model’s ability to learn data distribution. The MSE loss function has the characteristic of being twice continuously differentiable, which makes it perform well in the optimization process and facilitates the use of optimization algorithms such as gradient descent for model training. The loss function is formulated as:
\begin{equation}
	\begin{aligned}
		\mathbf{Loss} & =\mathbf{MSE} (\mathbf{r}, \mathbf{\theta})\\
		& = \frac{1}{N}\sum_n\sqrt{(r_n-\hat{r}_n)^2+(\theta_n-\hat{\theta}_n)^2},
	\end{aligned}
\end{equation}
where $r_n$ and $\theta_n$ are the real distance and angle, $\hat{r}$ and $\hat{\theta}$ denote the predicted counterparts.
\begin{algorithm}[tp]
	\caption{Transformer-Based Near-Field Beam Training Process}
	\begin{algorithmic}[1]
		\State \textbf{Input:} Predefined codebook $\mathcal{C}$, dataset $\mathcal{D} = \{(\mathcal{E}_n, r_n, \theta_n)\}_{n=1}^{N}$, $\mathcal{E}_n$ is the beam scanned map, learning rate $\eta$, batch size $B$, epochs $E$.
		\State \textbf{Output:} Trained model weights $\mathbf{\Omega}$.
		\State Initialize model parameters and optimizer state: $\mathbf{\Omega}$, $\eta$, and moments for Adam;
		\For{epoch $= 1$ to $E$}
		\State Shuffle dataset $\mathcal{D}$;
		\For{batch $= 1$ to $N/B$}
		\State Sample mini-batch $\{(\mathcal{E}_j, r_j, \theta_j)\}_{j=1}^{B}$ from $\mathcal{D}$;
		\State Compute $(\mathbf{\hat{r}}, \mathbf{\hat{\theta}})=\text{Transformer}(\mathcal{E}_j; \mathbf{\Omega})_{j=1}^B$;
		\State Compute loss: $\mathcal{L} = \frac{1}{B} \sum_{j=1}^{B} \text{MSE}(\hat{r}_j, \hat{\theta}_j, r_j, \theta_j)$;
		\State Compute gradients $\nabla_\mathbf{\Omega} \mathcal{L}$;
		\State \textbf{Update Model Parameters:}
		\For{each parameter $\omega \in \mathbf{\Omega}$}
		\State Compute parameter update: $\omega \gets \omega - \eta \frac{m_t}{\sqrt{v_t} + \epsilon}$;
		\EndFor
		\EndFor
		\EndFor
		\State \Return $\mathbf{\Omega}$
	\end{algorithmic}
	\label{algorithm2}
\end{algorithm}

\subsubsection{Beam training}
To implement the beam training process, we generate the channel in accordance with the procedure delineated in \textbf{Algorithm \ref{algorithm1}}. The comprehensive steps of the proposed Transformer-based near-field beam training algorithm are detailed in \textbf{Algorithm \ref{algorithm2}}. Training commences by initializing the model parameters and optimizer configurations, including the specification of the learning rate and momentum parameters for the Adam optimizer.

{The training process proceeds by iterating over a dataset consisting of uplink feedback signal strength maps and their corresponding ground-truth locations, repeatedly processed over a specified number of epochs.} At each epoch, the dataset is randomized to ensure statistical independence, and mini-batches are extracted. For each mini-batch, forward propagation through the Transformer model is executed to infer location predictions. The loss is computed by the MSE between the predicted and true locations. Subsequently, backward propagation computes gradients with respect to model parameters. These gradients are employed by the optimizer to update the model parameters, adjusting the learning rate and incorporating momentum to facilitate convergence. This iterative process persists over all epochs, refining the model to minimize the MSE loss. After training, the optimized model can be used to predict locations with accuracy. Ultimately, the trained model empowers the BS to scan the near-field codebook and directly infer the optimal codeword.

\subsection{Complexity Analysis}
In this section, we present a detailed analysis of the computational complexity associated with the proposed Transformer-based near-field beam training algorithm, focusing on both time complexity and memory consumption, which are quantified using floating-point operations (FLOPs) \cite{He2015CVPR}. The Transformer architecture consists of three convolutional layers, an encoder, and three fully connected layers. The time complexity of each convolutional layer is given by $ T_{\text{conv}} \sim \mathcal{O}\left( (F_h^{l-1} F_w^{l-1})(H W) C_{l-1} C_l \right)$, where $ F_h^{l-1} $ and $ F_w^{l-1} $ represent the input feature map dimensions, $ H $ and $ W $ are the kernel size, and $ C_{l-1} $, $ C_l $ are the input and output channels, respectively. The time complexity of the self-attention mechanism in the Transformer is expressed as $ T_{\text{attn}} \sim \mathcal{O}(n g^2 + n^2 g)$, where $ n^2 g $ accounts for the calculation of attention scores, and $ n g^2 $ is due to the computation of Query, Key, and Value matrices. For multi-head attention with $ h $ heads, the time complexity is $ T_{\text{MH}} \sim \mathcal{O}\left( n^2g/h \right)$, leading to an overall complexity of $ \mathcal{O}(n^2 g) $. The feed-forward network, consisting of two linear transformations and a non-linear activation function, has a time complexity of $ T_{\text{ffn}} \sim \mathcal{O}(n g g_{\text{ff}})$, where $ g_{\text{ff}} $ is the hidden layer size. The time complexity for a fully connected layer is $ T_{\text{fc}} \sim \mathcal{O}\left( (F_h^{l-1} F_w^{l-1}) C_{l-1} C_l \right)$. Therefore, the overall time complexity of a single Transformer layer is $T_{\text{Transformer}} \sim \mathcal{O}(n^2 g + n g g_{\text{ff}})$, where the self-attention mechanism dominates, resulting in a total time complexity of $ \mathcal{O}(n^2 g) $. To evaluate computational load, FLOPs serve as an additional metric. The FLOPs for a convolutional layer are $ \text{FLOPs}_{\text{conv}} = F_h F_w C_{\text{in}} C_{\text{out}} H_{\text{out}} W_{\text{out}}$, for a fully connected layer, $ \text{FLOPs}_{\text{fc}} = 2 C_{\text{in}} C_{\text{out}}$, and for self-attention, $ \text{FLOPs}_{\text{attn}} = 4 n^2 g$, where the factor of four accounts for the calculations of Query, Key, Value, and attention scores. Regarding memory consumption, the self-attention mechanism requires $ \mathcal{O}(n^2 g) $ memory to store attention weights for each pair of input tokens, while the feed-forward layers require $ \mathcal{O}(n g g_{\text{ff}}) $ memory allocation.
\begin{figure}[tp]
	\centering
	\includegraphics[scale=0.6]{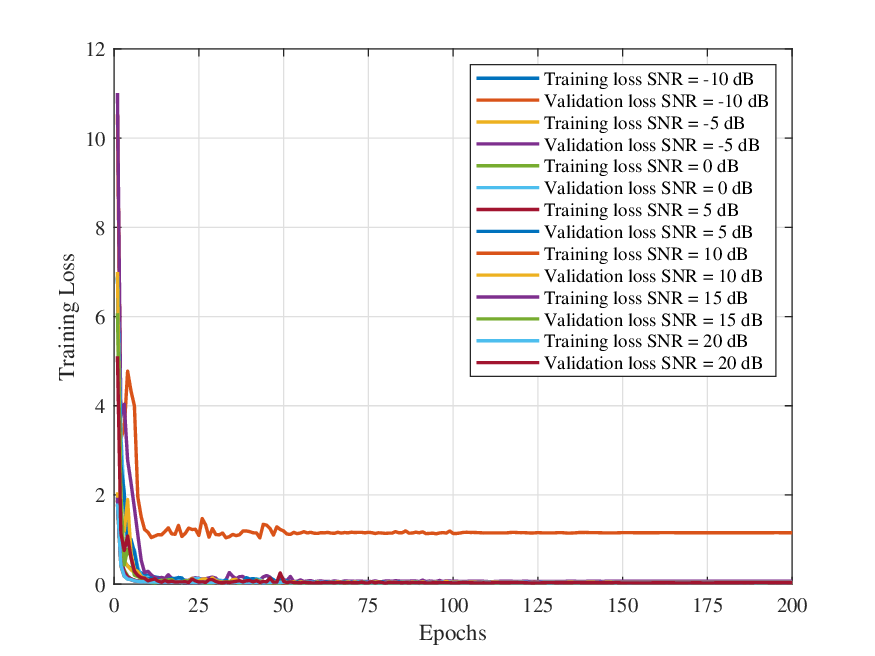}
	\caption{The variation curves of loss values on the validation set and training set with increasing number of trainings under different SNRs.}
	\label{TrainingLoss} 
\end{figure}

\section{Numerical Results}
\subsection{Dataset and Settings}
To assess the performance of the proposed beam training scheme, we performed experimental simulations utilizing the default configurations outlined in Table \ref{tab:simulationsetting}. These parameters are applicable to both the hierarchical beam training algorithm and the near-field channel model, as well as the second stage of beam training for hybrid cascaded channel models. The simulations focus on downlink beam training within an XL-MIMO system. The number of channel paths is established as $L = 4$. The channel gain for the LoS path is modeled as $k_0 \sim \mathcal{CN}(0, 1)$, and the non-line-of-sight (NLoS) paths are characterized by \( k_l \sim \mathcal{CN}(0, 0.001) \) for \( l = 1, 2, 3 \). The BS is equipped with \( N = 256 \) antennas and operates at a carrier frequency of $f_c = 60\ \text{GHz}$. The transmit power $P_t = 1W$, The angles of the device are randomly generated from a uniform distribution, $\phi \sim \mathcal{U}(-\pi/3, \pi/3) $, and the distances between the BS and the device are sampled from 3 meters to 100 meters. 

{For the details of data generation, the BS selects codewords to estimate the user's locations within the near-field region by utilizing uplink feedback. In the first stage, the BS conducts a search procedure employing four codewords. In the subsequent stage, two distinct uplink feedback radiation pattern datasets are generated, i.e., one through an exhaustive search encompassing all codes in the codebook, and the other by employing a reduced search space consisting of half of the codebook entries.} The near-field codebook comprises distance rings \( S=100 \) and angular diversity \( L=100 \), resulting in a total of \( C=L\times S=10,000 \) codewords for the second stage. The dataset employed for the training process consists of 700 samples in the initial stage and 7,000 samples in the subsequent stage, with 70\% designated for training, 10\% for validation, and 20\% for testing. The learning rate is initialized at 0.001 and reduced by 50\% every 50 epochs. To further evaluate the efficacy of the proposed algorithm, the two-step beam selection scheme is considered as two separate training tasks, and the accuracy is independently assessed through experiments with varying configurations on the two datasets.

\renewcommand{\arraystretch}{1.2} 
\begin{table}[tp]
	\centering
	\fontsize{8}{10}\selectfont
	\caption{The details of the experimental dataset}
	\label{tab:simulationsetting}
	\begin{tabular}{lccccc}
		\toprule
		\multirow{1}{*}{\textbf{Parameter}}&\multicolumn{5}{l}{\textbf{Assignment}}\\
		\hline 
		\multirow{1}{*}{Number of Transmit Antennas}&\multicolumn{5}{l}{N = 256} \\
		\multirow{1}{*}{Number of RIS Elements}&\multicolumn{5}{l}{M = 1024} \\
		\multirow{1}{*}{Signal Noise Ratio}&\multicolumn{5}{l}{[-10 dB, 20 dB], stride = 5 dB} \\
		\multirow{1}{*}{Sampling Distance Range}&\multicolumn{5}{l}{[3, 100]}\\
		\multirow{1}{*}{Sampling Angle Range}&\multicolumn{5}{l}{[-$\pi/3$, $\pi/3$]}\\
		\multirow{1}{*}{Number of Training Set for Each SNR}&\multicolumn{5}{l}{Stage 1: 70, Stage 2: 700}\\
		\multirow{1}{*}{Number of Validation Set for Each SNR}&\multicolumn{5}{l}{Stage 1: 10, Stage 2: 100}\\
		\multirow{1}{*}{Number of Test Set for Each SNR}&\multicolumn{5}{l}{Stage 1: 20, Stage 2: 200}\\
		\bottomrule
	\end{tabular}
\end{table}

\subsection{Beam Training Performance}
\begin{figure}[tp]
	\centering
	\includegraphics[scale=0.6]{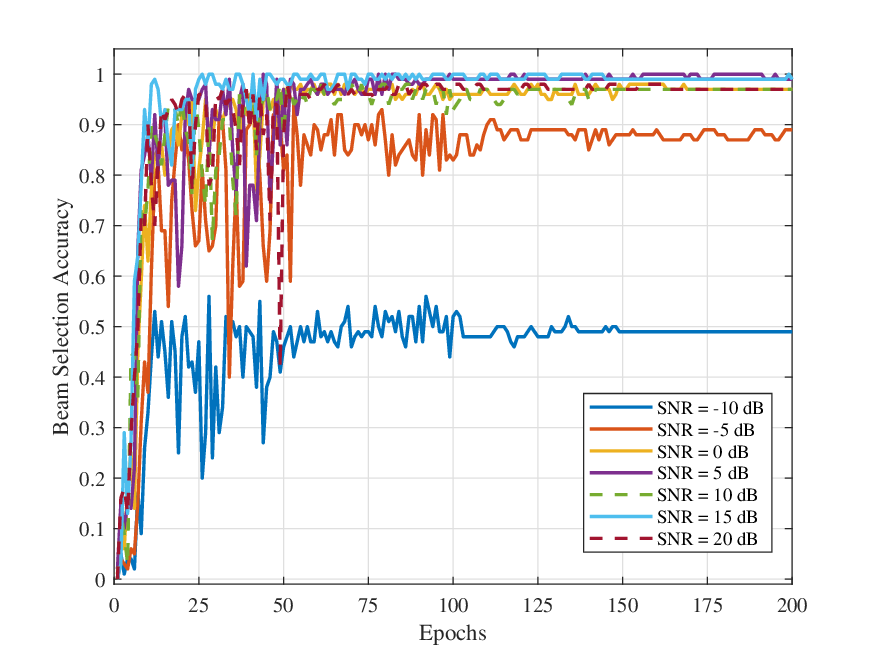}
	\caption{The variation curves of accuracy values on the validation set with increasing number of training epochs under different SNRs.}
	\label{TrainingAccuracy} 
\end{figure} 

\begin{figure}[tp]
	\centering
	\includegraphics[scale=0.6]{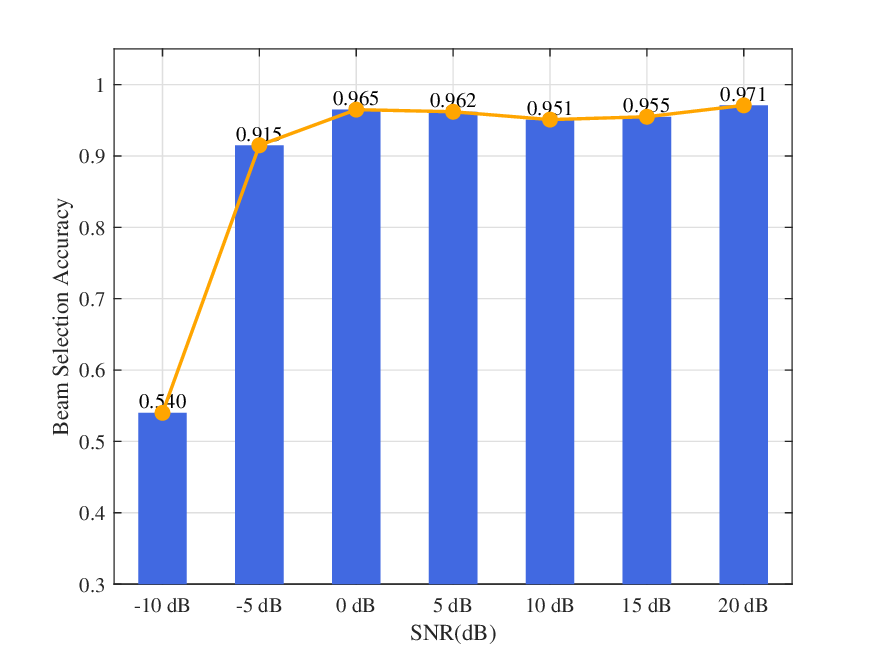}
	\caption{Beam prediction accuracy under different SNR condictions in the near-field by the proposed method.}
	\label{BeamPredict} 
\end{figure} 

\begin{figure}[tp]
	\centering
	\includegraphics[scale=0.58]{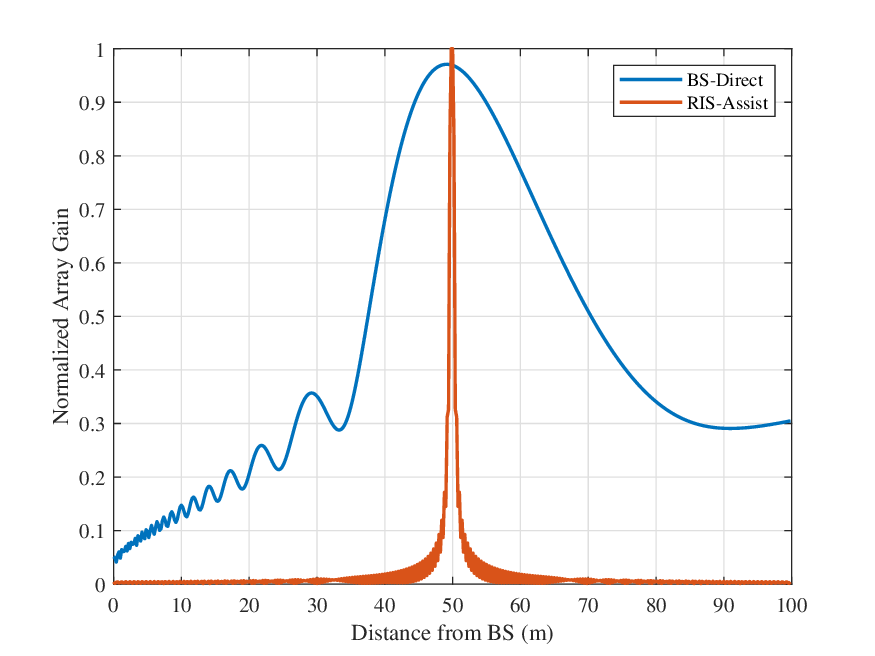}
	\caption{{Signal gain intensity variation at different distances from the BS in NFC.}}
	\label{risbs} 
\end{figure} 
{Fig. \ref{TrainingLoss} presents the training and validation loss curves under different SNR conditions, highlighting the influence of SNR on model performance.} The results indicate that higher SNR values lead to faster convergence and lower final loss. For SNRs of 20 dB and 15 dB, both training and validation losses rapidly approach zero after approximately 25 epochs, demonstrating robust convergence and minimal error in low-noise environments. At SNR = \{5 dB, 10 dB\}, while the final losses also reach around 0.01, the stability is less pronounced than at higher SNRs, suggesting that moderate noise levels increase the model’s error rate. Under lower SNR conditions, the validation loss stabilizes at around 0.05 for SNR = 0 dB, 0.13 for SNR = -5 dB, and exceeds 1.1 for SNR = -10 dB.

Fig. \ref{TrainingAccuracy} presents the training and validation accuracy curves under various SNR conditions, reflecting the impact of SNR on predictive performance. At SNR = 20 dB and SNR= 15 dB, the model’s accuracy quickly approaches 1.0 within the first 20 epochs. It remains stable, indicating excellent predictive capability in low-noise settings. At SNR = 10 dB and SNR = 5 dB, validation accuracy stabilizes at approximately 0.98 and 0.97, respectively, after about 30 epochs, demonstrating the model’s effectiveness even under moderate noise. However, as SNR decreases further, accuracy declines significantly. At SNR = 0 dB, accuracy settles around 0.96, and it hovers near 0.95 at SNR =-5 dB. At SNR = -10 dB, the training accuracy drops to 0.5.

Fig. \ref{BeamPredict} shows the beam prediction accuracy of the proposed method at different SNR conditions. At SNR = -10 dB, the accuracy is around 0.540, but improves to 0.915 at SNR = -5 dB. At SNR $\geq$ 0 dB, beam selection accuracy remains at 0.95, peaking at 0.971 at SNR = 20 dB. These results demonstrate that the proposed method performs robustly under high SNR conditions and excels in low-noise environments. {However, the performance degradation at SNR = -10 dB can be attributed to the fact that, under such low SNR conditions, the increased noise interference severely corrupts critical signal features. This leads to blurred spatial mapping of signals, which obscures fine-grained structural patterns essential for accurate beam selection. Moreover, the Transformer model’s reliance on precise pairwise correlations for its self-attention mechanism makes it particularly vulnerable to noise dominance, resulting in degraded decision-making accuracy.}

{Fig. \ref{risbs} illustrates the comparison of gain magnitude between the RIS-assisted beamfocusing scheme and the BS-directed communication across different distances. In details, when the user is located 50 meters from the BS, the 3 dB beamwidth extends up to 33 meters through direct communication approach. Whereas, the RIS-assisted mechanism can focus the beam onto a spot with a diameter of 1 meter.  Accordingly, the proposed RIS-assisted method significantly enhances signal quality, reduces interference, and improves overall system capacity.  By mitigating these interference effects through advanced phase control and beam optimization techniques, the approach shows considerable potential for overcoming the challenges posed by dense communication environments and enabling high-capacity communication.}

\begin{figure}[tp]
	\centering
	\includegraphics[scale=0.6]{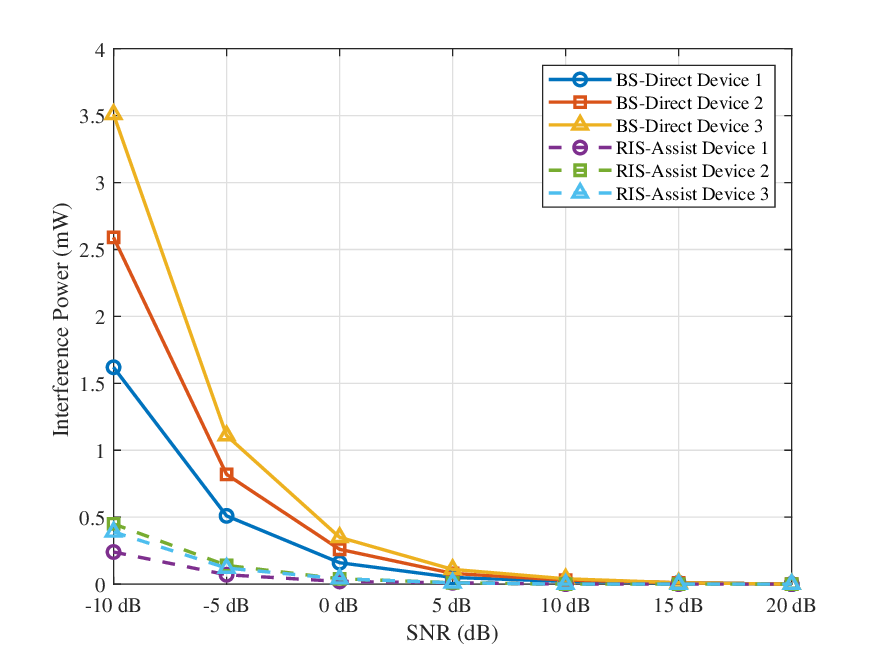}
	\caption{Interference power comparison for BS-directed \textit{vs.} RIS-assisted beamfocusing under varying SNR conditions.}
	\label{InterferencePower} 
\end{figure} 

\begin{figure}[tp]
	\centering
	\includegraphics[scale=0.6]{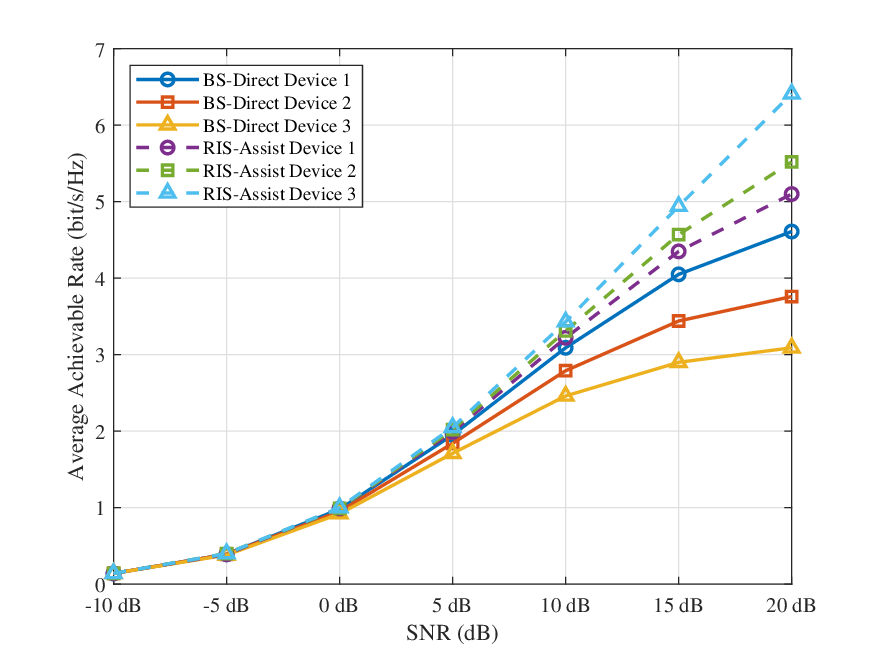}
	\caption{Average achievable rate comparison for BS-directed \textit{vs.} RIS-assisted beamfocusing under varying SNR conditions.}
	\label{AverageRate} 
\end{figure}

As shown in Fig. \ref{InterferencePower}, three devices are tested with respective distance and angle parameters: device 1 at $(5, -\pi/6)$, device 2 at $(15, \pi/6)$, and device 3 at $(25, \pi/8)$. The experimental results provide a comprehensive performance comparison between BS-directed and RIS-assisted communication, focusing on device interference and data rates under various SNR conditions. In the BS-directed approach with SNR = -10 dB, interference values for devices 1, 2, and 3 are 1.62 mW, 2.59 mW, and 3.51 mW, respectively. As SNR increases, these interference levels decrease, dropping to 0.16 mW, 0.26 mW, and 0.35 mW at SNR = 0 dB, and approaching 0 at SNR = 20 dB. These results indicate that while traditional communication confronts substantial interference at low SNR, improving SNR can mitigate this issue. In contrast, the RIS-assisted approach significantly reduces interference even at SNR = -10 dB, with values of 0.24 mW, 0.45 mW, and 0.39 mW for devices 1, 2, and 3, respectively{, due to near-field beamfocusing and interference nulling. By dynamically adjusting phase shifts, RIS narrows the beam coverage and suppresses interference toward unintended users, thereby ensuring more reliable communication.} When SNR $\geq$ 10 dB, RIS effectively eliminates interference, demonstrating strong anti-interference capabilities in multipath and challenging channel conditions.
\begin{figure}[tp]
	\centering
	\includegraphics[scale=0.58]{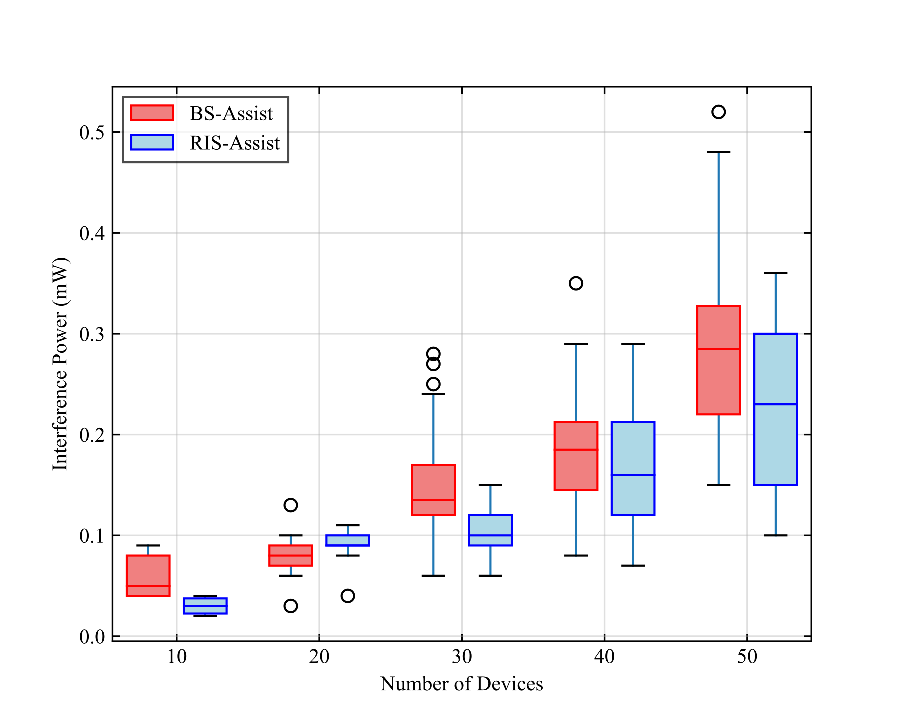}
	\caption{Interference power comparison for between BS-directed \textit{vs.} RIS-assisted based approach under different number of devices in the near field.}
	\label{InferenceNum}
\end{figure}

Fig. \ref{AverageRate} further compares the average achievable rates achieved by BS-directed and RIS-assisted communication under different SNR conditions. In the BS-directed scenario at SNR = -10 dB, all the three devices achieve a uniform rate of 0.14 bits/s/Hz. As SNR increases, these communication rates gradually increase, reaching 3.09 bits/s/Hz, 2.79 bits/s/Hz, and 2.46 bits/s/Hz at SNR = 10 dB and 4.61 bits/s/Hz, 3.76 bits/s/Hz, and 3.09 bits/s/Hz at SNR = 20 dB. Although higher SNR improves data rates, the increase is relatively slow at lower SNR levels. Comparatively, RIS-assisted communication consistently outperforms BS-directed communication in terms of data rates. At SNR = -10 dB, the rates are comparable to BS-directed, but when SNR $\geq$ 10 dB, RIS significantly enhances rates to 3.22 bits/s/Hz, 3.31 bits/s/Hz, and 3.43 bits/s/Hz at SNR = 10 dB, and 5.10 bits/s/Hz, 5.52 bits/s/Hz, and 6.41 bits/s/Hz at SNR = 20 dB. These results highlight RIS’s ability to improve channel conditions and optimize transmission efficiency. The experimental result indicate that RIS-assisted communication excels at reducing the device interference and enhancing data rates, especially in low SNR environments. As the SNR increases, the advantages of RIS become more pronounced, underscoring its potential in future wireless communication systems. By optimizing signal propagation and mitigating multipath effects, RIS emerges as a promising solution for substantially improving communication system performance.

Fig. \ref{InferenceNum} presents a comparison between RIS-assisted and BS-directed communication approaches across varying device densities, emphasizing notable differences in interference power management. The results highlight the advantages of RIS-assisted communication, particularly in high-density scenarios. As the number of devices increases, the RIS-supported method consistently demonstrates lower median interference power and reduced variability than the BS-directed approach. When there are 10 or 20 devices, RIS-assisted approaches maintain stable, low interference power levels, while BS-directed approaches exhibit more variable interference levels. As the device count rises to 30, 40, and 50, the RIS-assisted approach continues to effectively manage interference, with only a modest increase in median interference power. In contrast, the BS-directed method experiences a substantial rise in interference power, characterized by significant variability and numerous high outliers, especially at the larger number of device. In light of these findings, it is evident that RIS-assisted communication is capable of managing interference in densely populated environments and is an excellent solution for future communication systems that require high levels of stability and quality.

\begin{figure}[tp]
	\centering
	\includegraphics[scale=0.6]{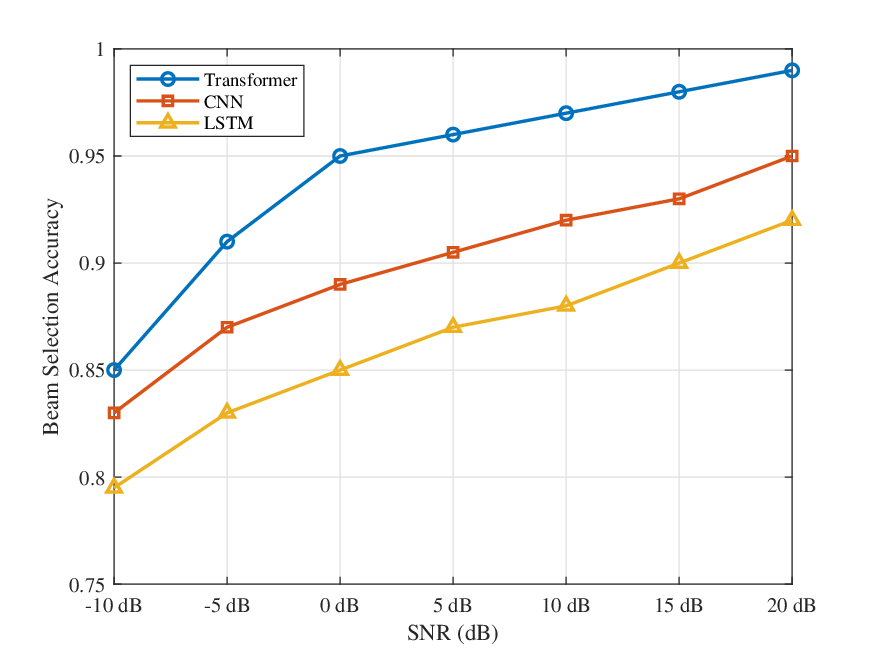}
	\caption{Comparison of beam prediction accuracy across the proposed method and baselines at varying SNRs in the first phase.}
	\label{firstStage}
\end{figure}

\begin{figure}[tp]
	\centering
	\includegraphics[scale=0.6]{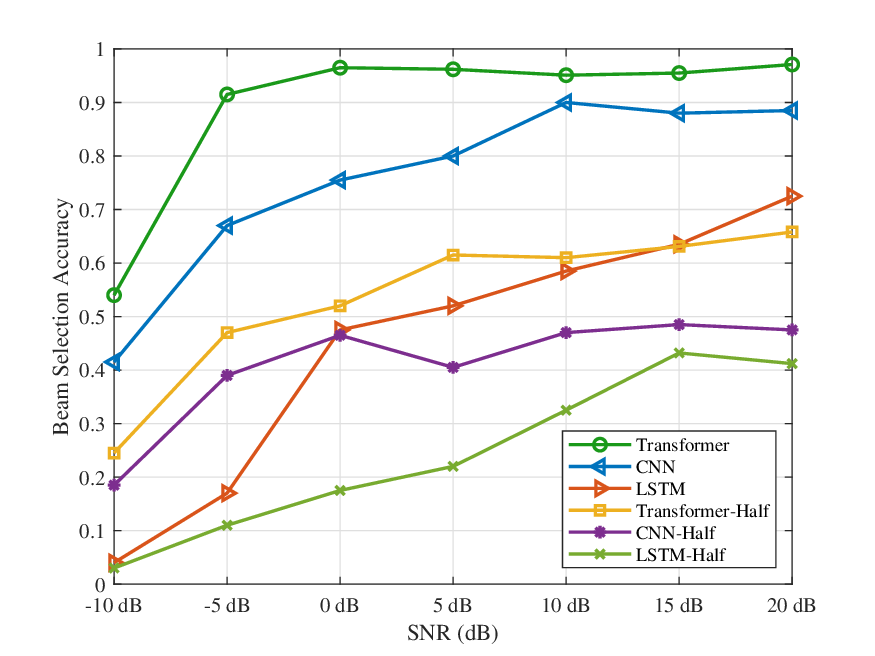}
	\caption{Comparison of beam prediction accuracy across the proposed method and baselines at varying SNRs in the second phase.}
	\label{compare}
\end{figure}
\subsection{Comparison with Baseline Methods}
Additionally, we implement experimental comparison with traditional algorithm, which are described as follow:
\begin{itemize}
	\item{\textbf{CNN} \cite{BTNFC4}}: A DL-based near-field beam training scheme with supplemental codewords, wherein the supplemental codewords are selected to perform beam training based on probability vectors obtained in the near-field beam training scheme.
	\item{\textbf{LSTM} \cite{Lim2021Tcom}}: The LSTM-based near-field beam training scheme modifies the beam search task into a beam selection problem, by leveraging the spatio-temporal correlation. The optimal beam selection is realized based on the cross-entropy loss function.
\end{itemize}

Fig. \ref{firstStage} presents a detailed comparison of baseline algorithms in the first stage, for predicting device location accuracy across a range of SNR values from -10 dB to 20 dB. Over this SNR range, all models show significant accuracy improvements as SNR increases. Specifically, at SNR $\leq$ 0 dB, the Transformer consistently maintains a robust accuracy of 85.1\%, significantly outperforming both the CNN and LSTM models. The CNN performs well at SNR $\geq$ 10 dB, achieving a peak accuracy of 95\% at SNR = 20 dB. However, it struggles to capture complex, non-local dependencies when SNR $\leq$ 0 dB. Despite its strength in modeling sequential data, the LSTM performs less effectively in NFC scenarios due to its limited ability to capture spatial complexity, resulting in a lower accuracy of 79.1\% at SNR = -10 dB. The Transformer model's superior performance can be attributed to its ability to dynamically focus on the most relevant parts of the input using global attention mechanisms. This enables the model to capture long-range dependencies in both spatial and temporal dimensions, which is crucial for beam prediction tasks in NFC. In low SNR conditions, the Transformer's attention mechanism effectively isolates signal-related features while suppressing irrelevant or noisy information, thereby maintaining high prediction accuracy even when signal quality is severely degraded. The Transformer-based approach consistently demonstrates advantages in beam prediction accuracy across a wide range of SNR conditions, especially in challenging low SNR environments. Its robust performance highlights the potential of Transformer architectures to enhance beam prediction in next-generation wireless communication systems, surpassing CNNs and LSTM.

To validate the capability of the proposed algorithm to reduce the search overhead, we perform the beam search based on only half of the codewords in the codebook, i.e., the generated beam mapping map has only half of the uplink feedback signal strength values. We generate the same amount of data to compare the beam training accuracy of their corresponding half-code map. Fig. \ref{compare} presents a detailed comparison of various algorithms in the second stage, including the Transformer-based method, CNN, LSTM, and the half-code map, in predicting the device location accuracy across all SNRs. The proposed Transformer-based algorithm outperforms all other models across the entire range of SNR conditions. It achieves superior accuracy in high SNR scenarios and maintains beam selection accuracy close to 97.2\% at SNR = 0 dB. This exceptional performance is attributed to the Transformer's ability to analyze complex data patterns and dependencies, enhancing its robustness against noise and interference. The Transformer's attention mechanisms enable it to focus on relevant features and suppress less informative ones, resulting in high accuracy across a wide range of SNRs. In comparison, the CNN performs well at higher SNR levels, reaching 90\% accuracy at SNR = 10 dB. However, when SNR $\leq$ 0 dB, the CNN exhibits increased variability and decreased accuracy due to its limited ability to handle noise and signal degradation in low SNR environments. The LSTM shows moderate performance, with accuracy improving gradually as SNR increases. However, it consistently underperforms relative to both the Transformer and CNN models at higher SNR environments. 

As shown in Fig. \ref{compare}, the reduced-beam search data of the models, particularly Transformer-Half, retain some benefits of the full model but suffer from lower accuracy, particularly in low SNR environments. The reduced beam searching map limits the model's ability to generalize under challenging conditions. Besides, CNN-Half shows minimal performance improvement across various SNR conditions, indicating that lower model complexity hampers its ability to capture diverse data patterns. These results highlight the importance of model complexity for accurately handling complex spatial relationships in beamforming tasks. At SNR 0 $\leq$ dB, the Transformer outperforms the other models due to its ability to capture long-range dependencies and focus on relevant features through its global attention mechanism. For the CNN, which relies on space feature extraction, struggle with signal degradation and noise, leading to undesirable accuracy. Similarly, the sequential characteristic of LSTM limits its ability to model the complex spatial relationships necessary for beam selection, resulting in awful performance under low SNR conditions.

\section{Conclusion}
{This paper proposed a RIS-assisted two-stage beam training framework aimed at enhancing beam selection performance across diverse near-field communication scenarios. Specifically, the conventional beam training problem was reformulated into a beam detection task, and an end-to-end Transformer-based beam selection scheme was developed to exploit spatial features in both angular and distance domains. Extensive simulation results validated the effectiveness of the proposed RIS-assisted beamfocusing approach, achieving a beam selection accuracy of up to 93\% and delivering an average performance gain ranging from 10\% to 50\% compared with baseline methods under various SNR conditions. These findings underscore the potential of integrating RIS-assisted beamfocusing with an MSE-driven beam-search strategy to significantly improve beamforming precision and efficiency in near-field IoT communication systems.}

\end{document}